\documentclass[useAMS,usenatbib]{mn2e}
\usepackage[]{graphicx}
\usepackage[]{epsfig}

\title[Characterization of the sodium layer at Cerro Pach\'on]
  {Characterization of the sodium layer at Cerro Pach\'on, and impact on laser guide star performance}
\author[B. Neichel et al.]
  {B.~Neichel,$^1$ C.~D'Orgeville,$^2$ J.~Callingham,$^3$ F.~Rigaut,$^2$ C.~Winge$^1$ and G.~Trancho$^4$ \\
  $^1$Gemini Observatory, c/o AURA, Casilla 603 La Serena, Chile\\
  $^2$Australian National University Research School of Astronomy and Astrophysics, Mount Stromlo Observatory, Cotter Road,\\ Weston ACT 2611, Australia\\
  $^3$Sydney Institute for Astronomy (SIfA), School of Physics, The University of Sydney, NSW 2006, Australia\\
  $^4$Giant Magellan Telescope Organization Corporation, PO box 90933, Pasadena, CA, 91109-0933, USA.}
\date{Released 2012 Xxxxx XX}

\pagerange{\pageref{firstpage}--\pageref{lastpage}} \pubyear{2012}

\def\LaTeX{L\kern-.36em\raise.3ex\hbox{a}\kern-.15em
    T\kern-.1667em\lower.7ex\hbox{E}\kern-.125emX}

\begin{document}

\label{firstpage}

\maketitle

\begin{abstract}
Detailed knowledge of the mesopheric sodium layer characteritics is crucial to estimate and optimize the performance of Laser Guide Star (LGS) assisted Adaptive Optics (AO) systems. In this paper, we present an analysis of two sets of data on the mesospheric sodium layer. The first set comes from a laser experiment that was carried out at Cerro Tololo to monitor the abundance and altitude of the mesospheric sodium in 2001, during six runs covering a period of one year. This data is used to derive the mesospheric sodium column density, the sodium layer thickness and the temporal behavior of the sodium layer mean altitude. The second set of data was gathered during the first year of the Gemini MCAO System (GeMS) commissioning and operations. GeMS uses five LGS to measure and compensate for atmospheric distortions. Analysis of the LGS wavefront sensor data provides information about the sodium photon return and the spot elongation seen by the WFS. 
All these parameters show large variations on a yearly, nightly and hourly basis, affecting the LGS brightness, shape and mean altitude. The sodium photon return varies by a factor of three to four over a year, and can change by a factor of two over a night. In addition, the comparison of the photon returns obtained in 2001 with those measured a decade later using GeMS shows a significant difference in laser format efficiencies. We find that the temporal power spectrum of the sodium mean altitude follows a linear trend, in good agreement with the results reported by \citet{Pfrommer10B}. 
\end{abstract}


\section{Introduction}

\subsection{Mesospheric sodium layer}

Sodium (Na) atoms are believed to be deposited in the high atmosphere by meteoritic ablation. They form a layer
whose average altitude lies in the 90-95 km altitude range above sea level. This altitude can vary from 85 km to 105 km depending on the particular physical conditions. The sodium abundance displays typical seasonal variations of a factor of 2 to 4, with the minimum and maximum seasonal abundance occurring in summer and winter, respectively \citep{Moussaoui10A}. Significant abundance variations on hourly, daily and yearly time scales have been reported, even on time scales of a few seconds to a few hours \citep{Clemesha95,Miloni98,Sullivan00,Butler00,Michaille01}.  The sodium layer mean altitude and width also exhibit significant variations on short and long time scales: the layer is in average lower in altitude (by 1 to 2 km) and thinner (by 1 km) at the equinoxes, and higher in altitude and thicker at the solstices \citep{Papen96}. Most of the short time scale abundance and mean altitude variations can be traced down to the appearance of relatively short-lived, high density, thinner layers within the main sodium layer, which are refered to as sporadics \citep{Clemesha96}.

\subsection{Laser Guide Stars in Adaptive Optics}

Observations with Adaptive Optics (AO) are limited to certain areas on the sky due to the requirement of a stellar source to measure the wavefront distortions. A solution to this sky coverage problem is to create an artificial guide star with a laser \citep{Foy85}.  A so-called ``Sodium LGS'' can be created from resonant backscattering of mesospheric sodium atoms. Today, almost all leading telescopes are equipped with LGS-AO systems \citep{Amico2010}, and recently, the Gemini Observatory produced the first sodium LGS constellation to feed a Multi-Conjugate Adaptive Optics (MCAO) system at the Gemini South 8-meter telescope.
With the increasing prevalence of such LGS systems, it is becoming more critical to understand and investigate the intrinsic properties and characteristics of the mesospheric sodium layer. For instance, a good knowledge of the sodium abundance is crucial to understand the fluctuations of the brightness of the LGS during the year, and hence the impact and limitations on the AO performance. This information is also helpful to optimize queue observations. Also of importance is the study of the sodium layer interaction with a given laser format so this interaction can be optimized, as different laser formats can lead to very different results \citep{Rochester12,Holzlohner10}.

\subsection{Sodium parameters for Laser Guide Stars}

The following characteristics of the sodium layer are key for the AO performance:

\begin{itemize}
\item Mesospheric sodium column density.
\item Temporal behavior of the sodium layer mean altitude.
\item Sodium layer thickness.
\end{itemize}

\noindent
\textit{Mesospheric sodium column density}\\
The mesospheric sodium column density is the most crucial parameter to characterise so that the effectiveness of the sodium LGS technique can be maximised. This is because, in absence of saturation or at low saturation levels typically achieved with currently existing LGS AO systems, the required laser power per beacon is directly proportional to the sodium column density. Therefore, the measurement of the minimum sodium column density over a year sets an upper limit for laser power requirements. Additionally, the seasonal and nightly statistics of sodium density fluctuations, which are equivalent to LGS magnitude fluctuations, provide insights into the expected long and short term system performance such that the efficiency and flexibility of queue scheduling can be maximised.\\

\noindent
\textit{Temporal behavior of the sodium layer mean altitude}\\
The rate of mean sodium altitude variations is of prime importance for optimum removal of the focus mode in AO-corrected science images. Focus adjustments in the LGS path of AO systems must distinguish
atmosphere-induced focus terms from slow drifts of the guide star altitude. The faster the mean
sodium altitude varies, the more difficult it is to distinguish those two effects and adequately correct
for them. GeMS is equipped with a Slow Focus Sensor (SFS) that monitors the sodium altitude drifts. Knowledge of the required rate at which this sensor should run is thus very important. On the other hand, for AO systems working with a LGS constellation like GeMS, the differential focus error between the LGS is also of importance. A differential focus between the LGS would produce a signal that could not be properly treated by a tomographic reconstructor, and would therefore lead to non-negligeable loss of performance out of the AO system. The temporal behavior of this differential error is key for the design of future AO systems on Extremely Large Telescopes \citep{Pfrommer2012,Herriot12, Diolaiti12}. \\

\noindent
\textit{Sodium layer thickness}\\
The sodium thickness will determine the LGS spot elongation as viewed from a subaperture located near the edge of the telescope pupil (this is $\sim$1 arcsec for an 8 m telescope and comparable to the LGS spot size).
The larger the telescope, the more elongated the spot, the more sensitive the adaptive optics system
is to noise and therefore the more laser power is required to achieve a given performance. 

\subsection{Sodium data}
With the increasing use of sodium LGS in astronomical AO, many studies have been done in the past years to characterize more precisely the properties of the sodium layer in the light of AO requirements. Models have been developed to predict the sodium abundance and expected photon return for different sites \citep{Moussaoui09,Holzlohner10}. The temporal behavior of sodium profiles, and the variations of the mean altitude have also been studied in detail \citep{Herriot06,Pfrommer10B}. However, whereas the sodium layer had been observed quite extensively at several locations, no measurements had been made yet in Chile at or near the latitude of Cerro Pach\'on. The Gemini Observatory therefore initiated a year-long sodium monitoring campaign at the Cerro Tololo Inter-American Observatory (CTIO) located only a few kilometers away from the Gemini South telescope. In 2001, during a series of 6 runs, a laser experiment was carried out at Cerro Tololo to monitor the abundance and altitude of the mesospheric sodium. The goal was to characterize yearly, nightly and hourly the mesospheric sodium variations at CTIO/Gemini latitude \citep{Celine03}.
This set of data is compared with the results obtained during the first year of the Gemini MCAO System (called GeMS) commissioning. GeMS started on-sky commissioning in January 2011, at a rate of 5 to 7 nights per month \citep{Rigaut12}. This first period of commissioning lasted 5 months, after which GeMS entered a shutdown phase for engineering upgrades. On-sky operations resumed in November 2011, and continued up to May 2012. During this first period of commissioning, we gathered data about the sodium return as seen by the LGS wavefront sensors, and the differential focus between the LGS. The differential focus error analysis will be presented in a forthcoming paper. \\

This paper is organized as follows. Section 2 presents the CTIO campaign along with the data reduction and results. Section 3 presents the GeMS data, data reduction and results. Section 4 compares the results of both data sets, and discusses the impact of the laser format on performance.

\section{The CTIO/Tololo campaign}

\subsection{Overview}
The experimental setup has been described in detail in \citet{Celine03}, here we only present its main characteristics. The experiment involved launching a laser beam whose wavelength was tuned to the sodium D2 absorption line (589 nm) to the sky such that mesospheric sodium atoms were excited to higher energy levels. Sodium atoms were excited with a low-power continuous-wave laser whose interaction with sodium atoms was relatively well-known \citep{Milonni99}. The laser equipment included a 6-7~W multi-line argon-ion laser pumping a commercial ring-dye laser. The laser output power was monitored in real time at a ~0.3 Hz rate so that sodium density fluctuations could be calibrated out from laser power fluctuations. The on-sky 589 nm launched power was in the $\sim$100-200 mW range and sent to the sky by a custom launch telescope. The laser beam, a truncated gaussian of $\sim$250~mm at the $1/e^2$ intensity points, was projected about 3 degrees off zenith at a fixed angle. The 0.9 m CTIO telescope and the University of Michigan Schmidt telescope were used to image the laser guide star thus created. They were respectively at a distance of 140 m and 110 m away from the laser projection system yielding a spot elongation of $\sim$35~arcsec. Images were taken with successive exposures of typically 10 seconds but this period of data acquisition varied depending on the photometry of the night. It reached up to 30 seconds on some of the observations. Standard stars were also observed a few times per night in order to derive atmospheric transmission. Provided that all due telescope calibrations were performed, and that the atmosphere optical transparency was measured during the night, the CCD data then contained all information necessary to retrieve the sodium column density, sodium layer width and relative altitude. The absolute altitude could also be derived by a triangulation method based on the natural guide star trails seen on the CCD behind the laser streak, however this method has not been implemented on the data yet, hence only relative altitudes will be presented here.

\subsection{Data and data reduction}
\label{ddr}
Observations were spread over one year from February 2001 to February 2002 to allow characterization of hourly,
nightly and yearly variations of the sodium layer parameters. Five runs of 7 to 10 nights were performed in February, May,
September, November 2001 and February 2002. Those particular months were chosen to match the expected minimum
(November/December) and maximum (May) of the sodium column density sinusoid-like variations. Table \ref{ttable} indicates the dates of each run and the resulting number of useful nights when data was taken.\\

\begin{table}
 \caption{Data summary. ``useable data'' means that the photometry was reliable enough for laser profiles not to depend on weather variations (e.g. variable cirrus cover).}
 \label{ttable}
 \begin{tabular}{@{}ccc}
  \hline
  Run number & Date & Night with useable\\
  &  & laser data \\
  \hline
  1 & Feb, 11-20, 2001 & 7 \\
  2 & May, 2-11, 2001 & 3 \\
  3 & Aug 31 - Sep 6, 2001 & 2\\
  4 & Nov 25 - Dec 1, 2001 & 6\\
  5 & Feb 23 - Mar 2, 2002 & 5\\
  \hline
 \end{tabular}
\end{table}

\begin{figure}
 \begin{center}
    \includegraphics[width = 0.9\linewidth]{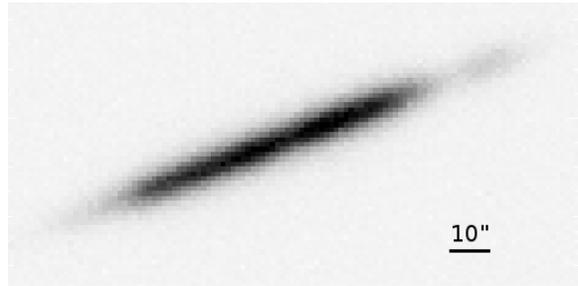} 
  \end{center}
  \caption[] {\label{strike} Example of reduced image obtained on the 0.9m telescope CCD, showing the laser streak. This image was taken at UT=00h51min on September 4th, 2001.}
\end{figure}

Data reduction follows the procedure/method described in \citet{Celine03}. The first part of the work consisted of retrieving, cleaning and organizing all the data from the compact disks (CDs) and notebooks the ten year-old data was stored on. All images were flat-fielded and bias-substracted using regular IRAF procedures. The nature of the flat fields varied from dome to sky flats depending on the run and photometry of the night. In one instance (Run 4) dark currents were also substracted. Fig.~\ref{strike} shows an example of the laser streak (data from May 10th 2001) once the image has been reduced. The next steps were to calibrate the flux received from the LGS into photons/s/cm$^2$/W and convert it into a sodium column density in atoms/cm$^2$. A first part of this calibration process consisted in deriving a zero point (ZP) and the atmospheric transmission for each night. This was based on observations of standard stars conducted during the night. These standard stars were observed either at the 0.9 m directly, or the nearby Schmidt telescope. Additionally, some standard stars were observed through the same Na filter, whereas other stars were observed with a $V$ filter. We derived the different factors to convert the flux into a final ZP based on the notes and data that we had.
The second part of this calibration concerns normalization by the laser flux. Laser power data for the last three runs (September 2001, November 2001 and February 2002) had been logged automatically which minimised any transcript errors. However, laser power data for the first two runs (February 2001 and May 2001) was only available as handwritten notes, therefore increasing data uncertainty and limiting the calibration accuracy. The conversion from sodium return measured in photons into sodium column density follows the same assumptions as in \citet{Celine03}, using an absorption cross section value of 1.0 10$^{11}$ cm$^2$ (see equation (1) in \citet{Celine03} and figure 2 in \citet{Miloni98}). Note that the elapsed time of 10 years between data taking and data reduction unfortunately results in some uncertainties regarding calibrations. This is especially true for Run 1 and 2 during which the data log was only available as handwritten notes. 
The errors introduced by the uncertainties in the calibration data are hard to quantify due to the different observation conditions at different points of the year. It is our estimate that such errors, in a worst case scenario, should be no larger than 25\%. We also expect the seasonal sodium abundance variations to have larger amplitudes than what can be accounted for by this error. Therefore, the error introduced by the uncertainties in the calibration procedure only adds to a noise floor that the sodium abundance variation will be clearly above, ensuring our proceeding analysis is sound. \\


\subsection{Results}

\subsubsection{Sodium profiles}
\begin{figure}
  \begin{center}
    \begin{tabular}{c}
     \includegraphics[width = 0.9\linewidth]{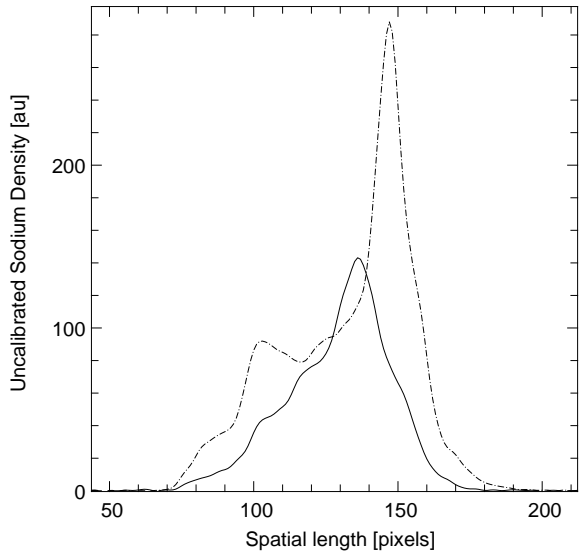}
    \end{tabular}
  \end{center}
  \caption[] { \label{profiles} Two sodium layer intensity profiles extracted from the same night: February 25th 2002,  UT=01h35min (solid line) and UT=07h45 (dashed line). Flux is expressed in Arbitrary Unit (AU). }
\end{figure}

By integrating over the image it is possible to derive an intensity profile that can be used to investigate such intrinsic properties of the sodium laser as equivalent width and mean altitude. Fig. \ref{profiles} shows two profiles that have been extracted from the 25th of February 2002 (run 5) at respectively UT=01h35min (solid line) and UT=07h45 (dashed line). The former profile demonstrates a typical laser profile shape, while the latter profile shows that a double asymmetric peak has developed, where the intensity of the second peak is almost twice stronger as the first. This example illustrates the variability that one can expect during a given night and emphasizes the importance of characterizing such behavior.


\begin{figure}
  \begin{center}
    \begin{tabular}{c}
     	\includegraphics[width = 0.66\linewidth]{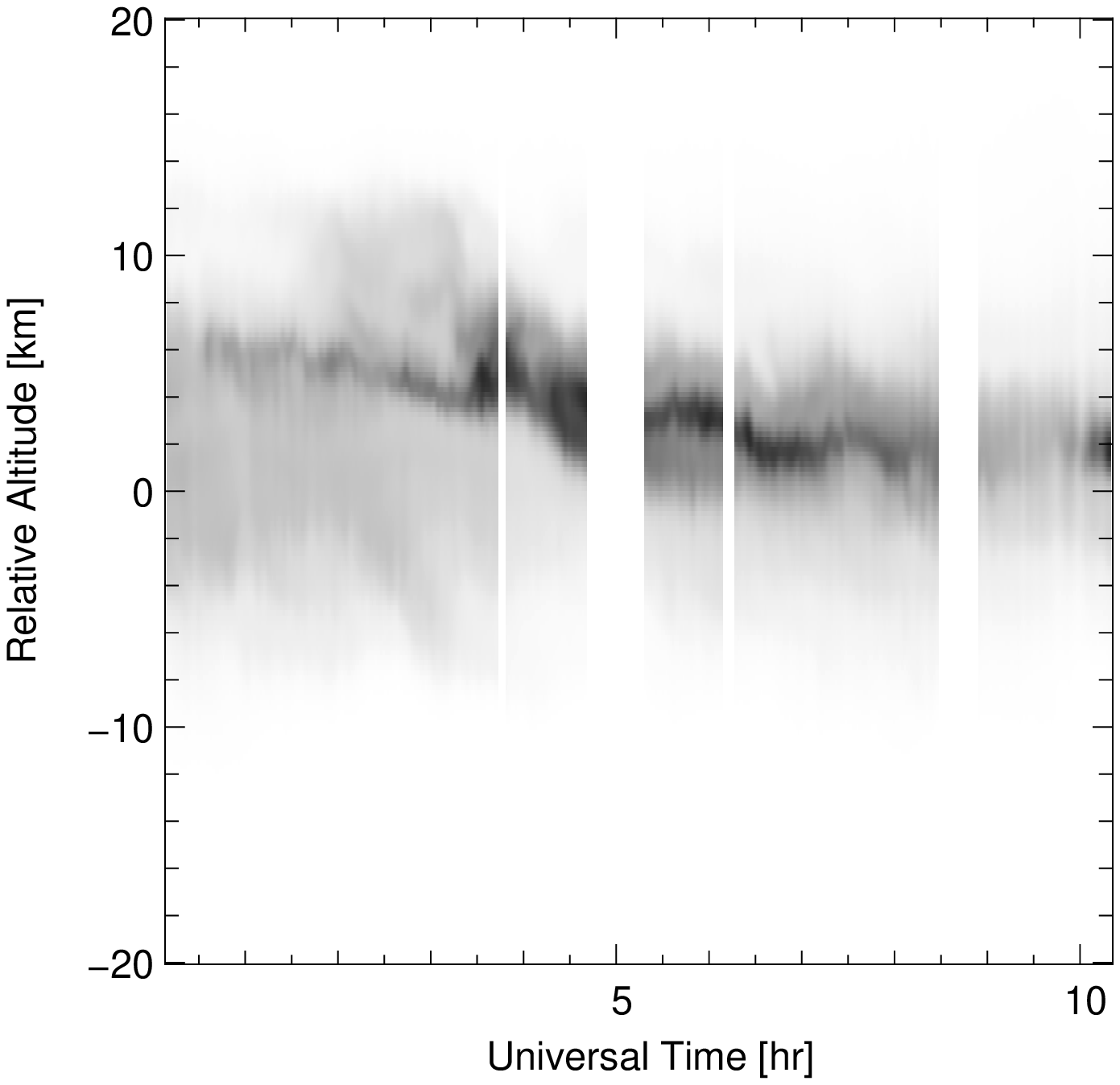}\\
	\includegraphics[width = 0.66\linewidth]{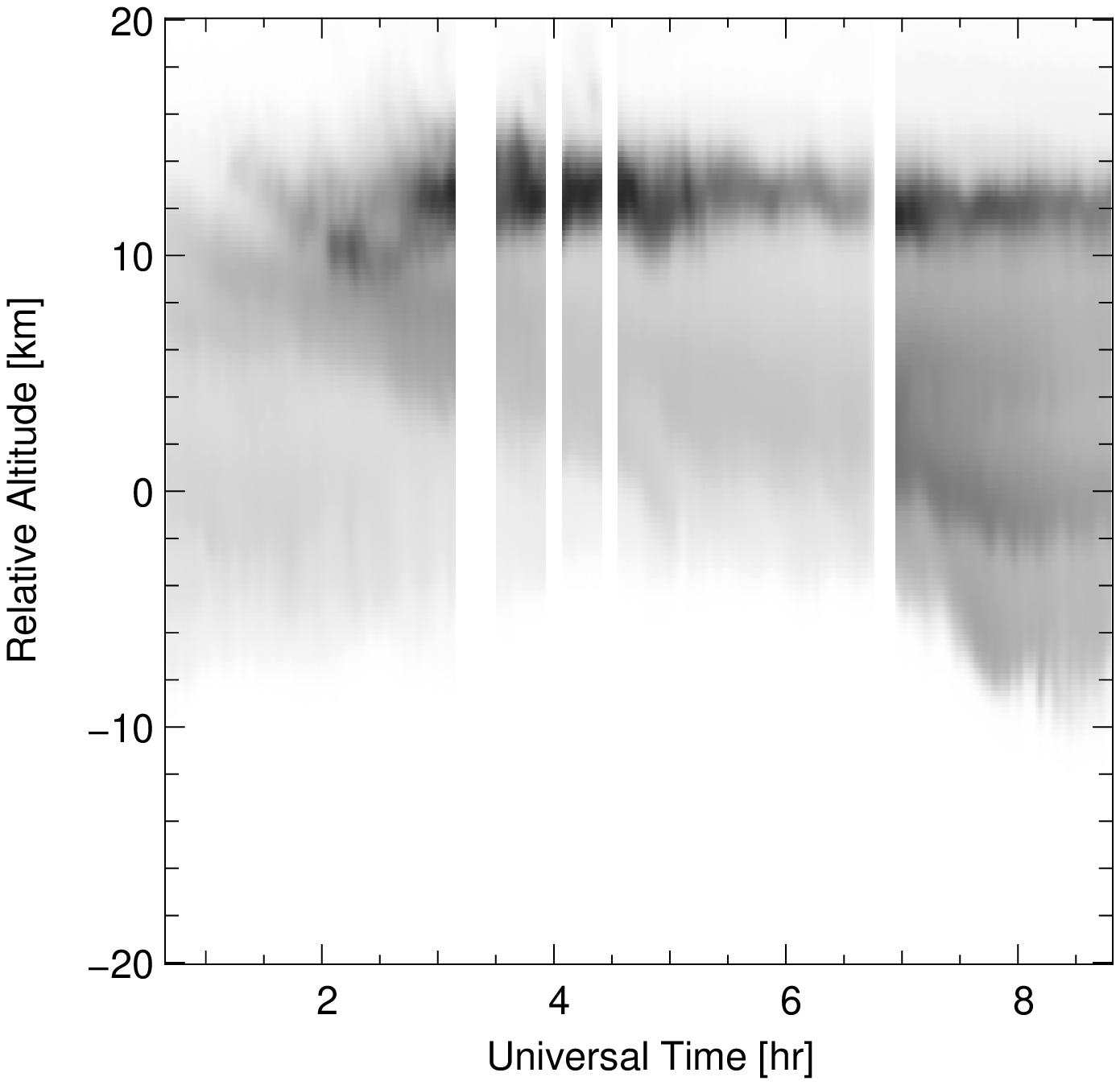} \\
     	\includegraphics[width = 0.66\linewidth]{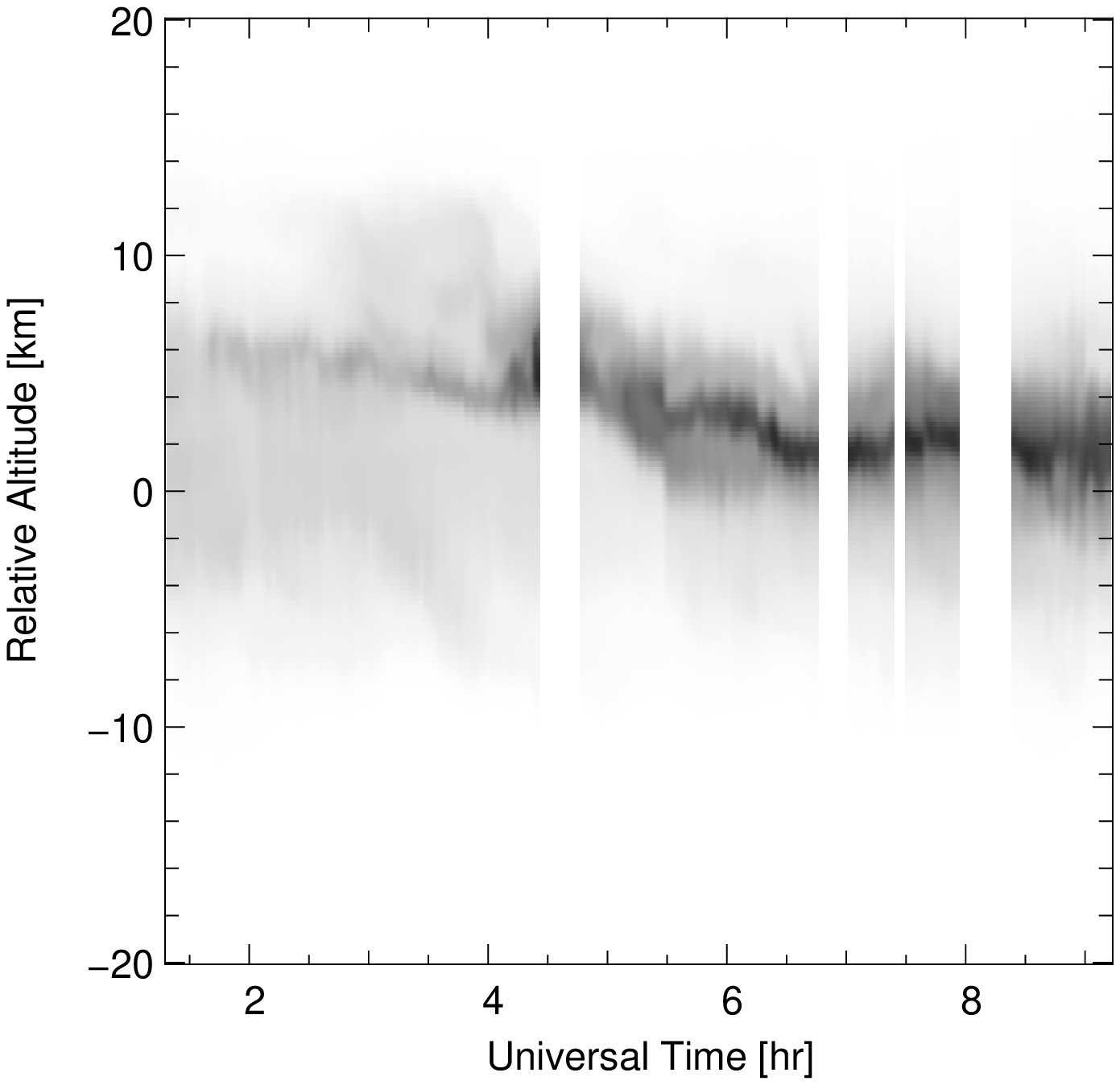}
    \end{tabular}
  \end{center}
  \caption[] { \label{profile} Sodium profiles for 3 sample nights. From top to bottom, the nights of: 4th September 2001, 27th November 2001, 25th February 2002. The white columns correspond to time intervals in the observation night when no data was taken because there was a problem with the telescope and/or laser.}
\end{figure}

\noindent
Concatenating all the different profiles, one can follow this evolution over a whole night. Profiles have been generated for all the nights which had data that was of high quality (see Table 1). Fig.~\ref{profile} shows three different sample nights. The previous example of February 25th is shown as the last plot at the bottom on Fig.~\ref{profile}. The event seen on the 2nd profile above appears around 05h UT and it develops up to the end of the night. The different examples in Fig.~\ref{profile} illustrate the high variability of the sodium layer in terms of profiles, as reported by \citet{Pfrommer10A,Pfrommer10B}. 

\subsubsection{Sodium equivalent width}
\begin{figure}
  \begin{center}
    \begin{tabular}{c}
	\includegraphics[width = 0.9\linewidth]{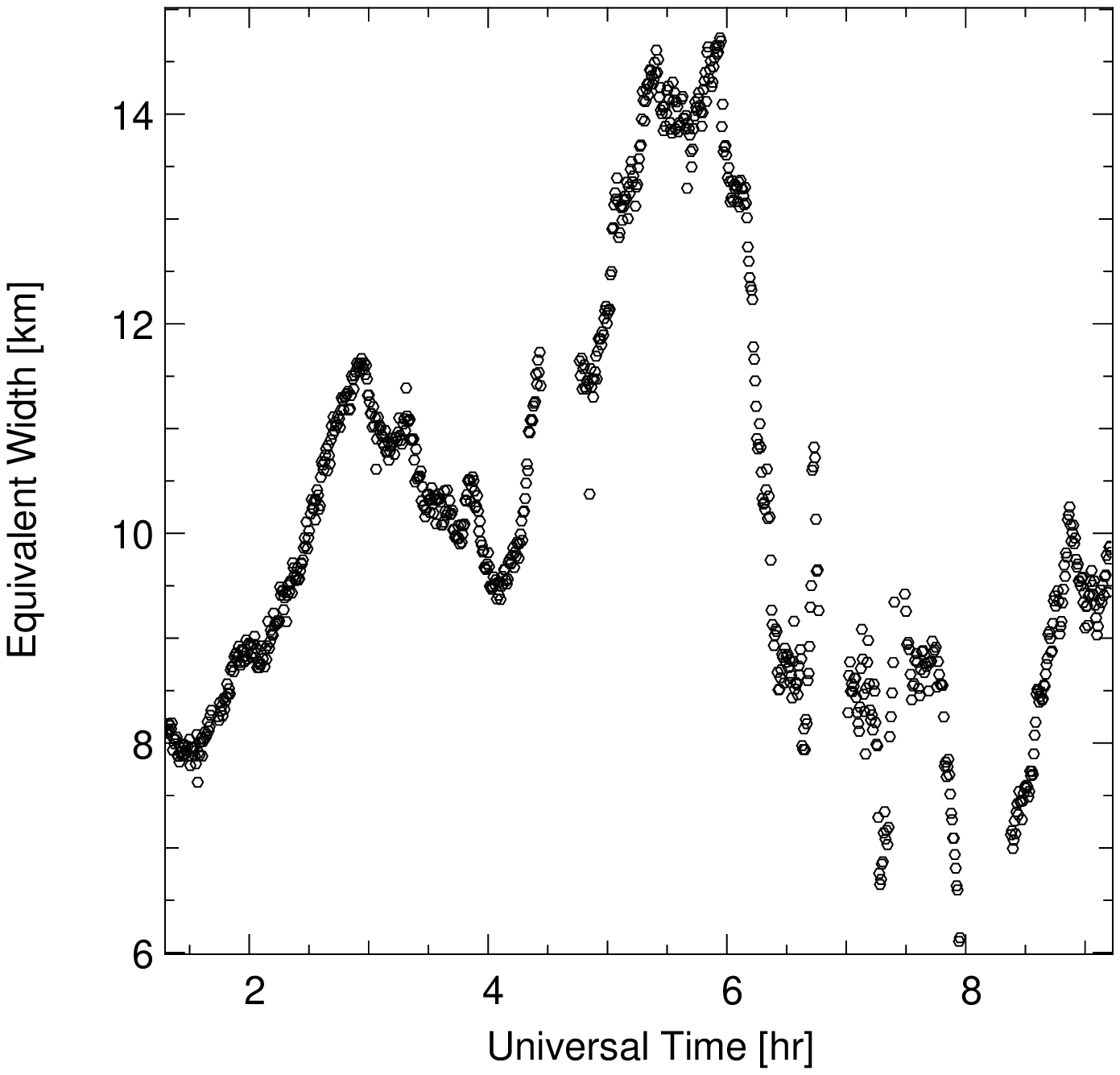} \\
       \includegraphics[width = 0.9\linewidth]{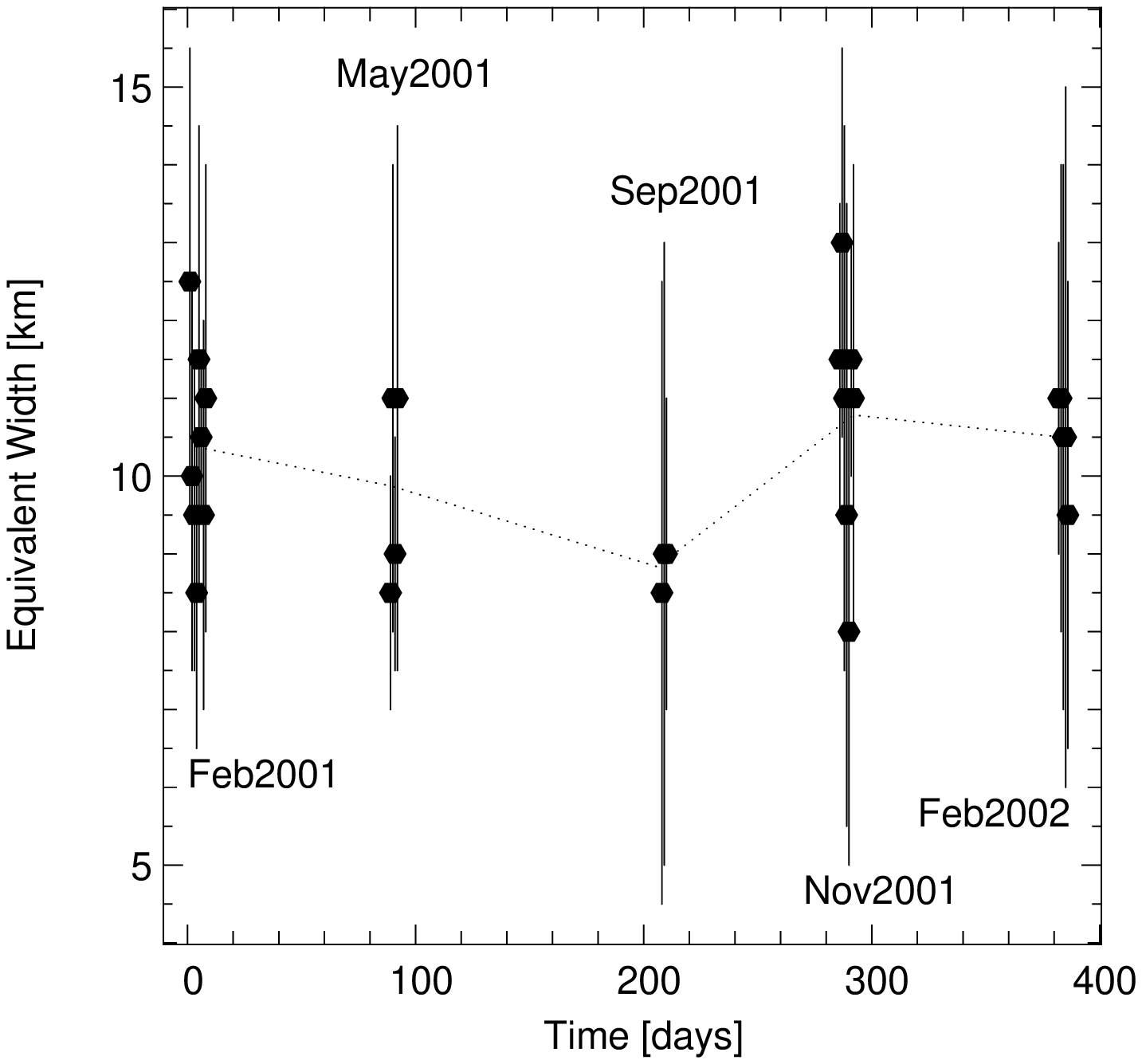}
    \end{tabular}
  \end{center}
  \caption[] { \label{eqw} Variation of the sodium profile equivalent width. Top is for the night of the 25th February 2002. Bottom is the average equivalent width per night for all the data-set. Error bars represent the minimum and maximum values per night.}
\end{figure}

For each profile, we estimate an equivalent width for the sodium thickness. For this, we first normalize each profile by its maximum. The equivalent width is then found by forming a rectangle with a height equal to one, and finding the width such that the area of the rectangle is equal to the area in the profile. For profiles with irregular shapes, the equivalent width is more robust than a Full Width at Half Max (FWHM) estimation. 
The equivalent width gives an estimate of the spot elongation to be expected on the Laser Guide Star WaveFront Sensor (LGSWFS). In the small angle approximation, the angular elongation $\gamma$ is given by:
\begin{equation}
 \gamma = \frac{L \Delta H}{H^2 + H\Delta H} \quad,
\end{equation}
where $\Delta H$ is the thickness of the sodium layer, H is the low altitude of the sodium layer, and L is the distance between the pupil subaperture and the laser launch telescope. The top plot of Fig.~\ref{eqw} shows the evolution of the equivalent width for the night of 25th February 2002. The bottom plot of Fig.~\ref{eqw} shows the average width per night for the entire set covering one year of data. Average equivalent widths are on the order of 10~km, as expected, which leads to a spot elongation of $\sim$1~arcsec for the Gemini telescope whith its on-axis launch configuration. Significant variations on hourly and daily basis are seen. The equivalent width can change by almost a factor of two within a given night. A small correlation with seasons can be detected: the sodium layer is somewhat thinner (resp. thicker) around September (resp. February) by $\sim$1~km. This was also observed at other latitudes as reported by \citet{Papen96}. The largest equivalent widths seen on this data are on the order of $\sim$15 km, which corresponds to an elongation of 1.3 arcsec for the Gemini telescope. To avoid clipping the LGS spots on the WFS subaperture, this maximal expected elongation should be used to set the requirements on the WFS subaperture Field Of View (FoV). The GeMS subaperture FoV is 2.8 arcsec, which allows for the spot elongation and the kernel due to seeing. 

\subsubsection{Sodium mean altitude}

As described in the introduction, variations in the mean altitude of the sodium layer will be interpreted by the LGS AO system as focus variations, deteriorating the performance of the AO correction. A variation of $\Delta h$ in the mean altitude of the sodium layer will cause an rms phase error (piston-removed) of:
\begin{equation}
\sigma_{\delta} = \frac{1}{16\sqrt{3}}\frac{D^2 sin(\psi)}{(h - h_0)^2}\Delta h \quad,
\label{eq_foc}
\end{equation}
where $D$ is the diameter of the telescope, $\psi$ is the zenith angle, $h$ is the mean sodium altitude and $h_0$ is the telescope altitude. Using Marechal's approximation, we can estimate the loss of performance in term of Strehl Ratio (SR) due to this defocus phase error as: 
\begin{equation}
\mbox{SR} = \mbox{exp}\left[-\left(\frac{2\pi\sigma_{\delta}}{\lambda}\right)^2\right]\quad,
\label{eq_foc2}
\end{equation}
where $\lambda$ is the observation wavelength.\\

\begin{figure}
  \begin{center}
    \begin{tabular}{c}
     \includegraphics[width = 0.9\linewidth]{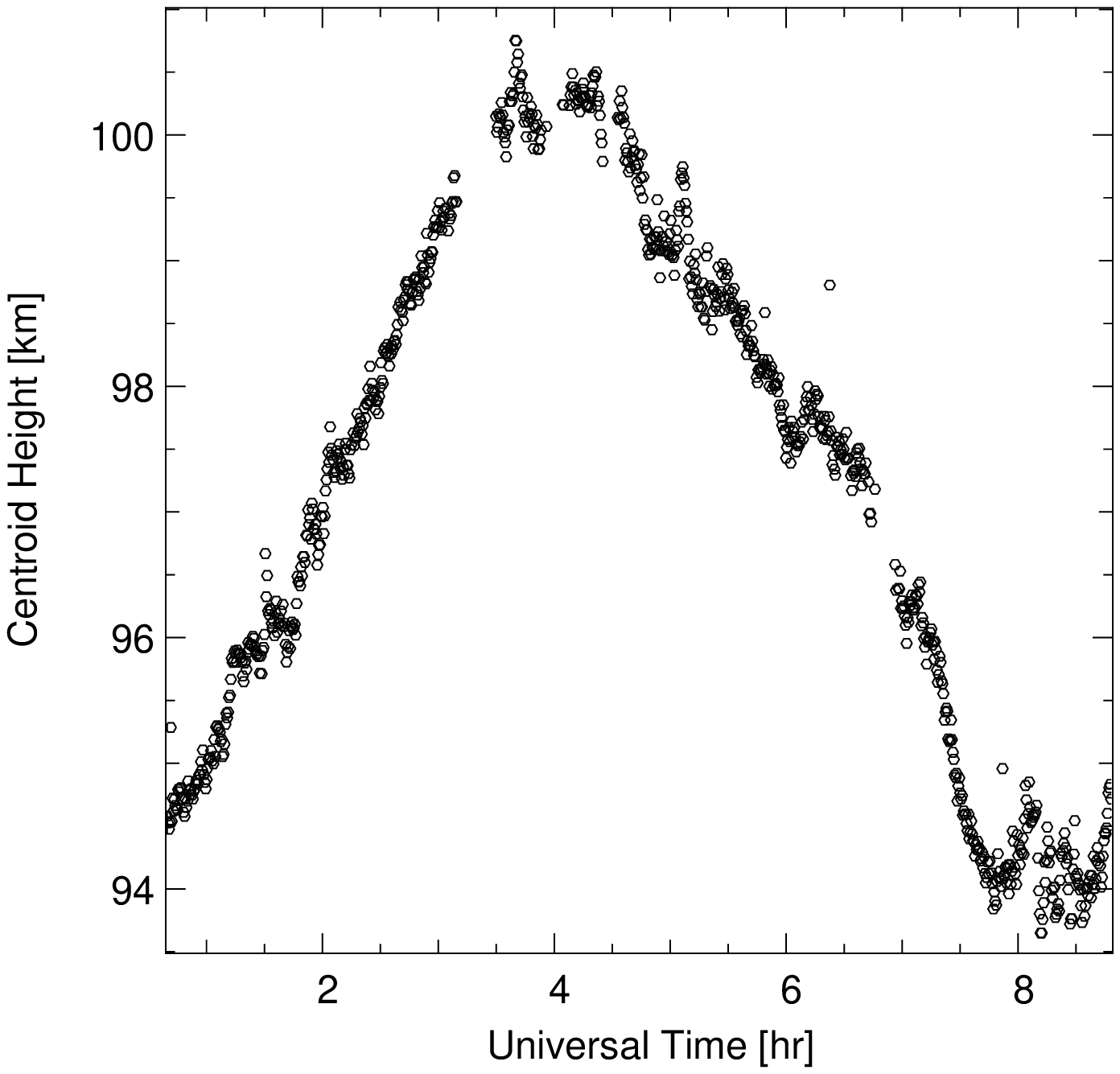} \\
      \includegraphics[width = 0.9\linewidth]{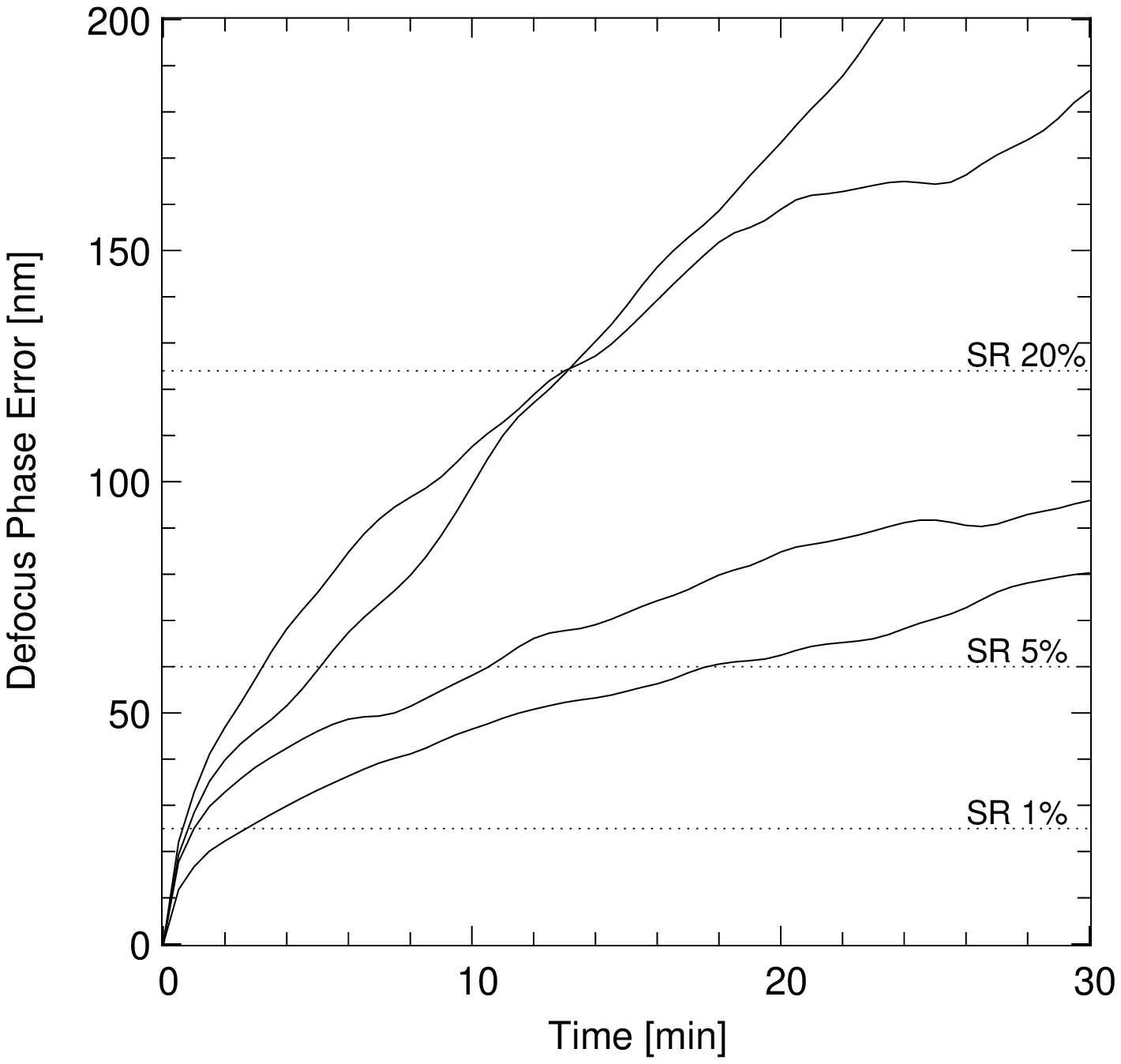} 
    \end{tabular}
  \end{center}
  \caption[] { \label{mean_altitude} Top: Variation of the mean altitude of the sodium layer for the night of 28th of November 2011. Bottom: Average of the mean altitude change for given $\Delta t$ in minutes. Four nights are illustrated here, from top to bottom: 28th November 2001, 27th November 2001, 4th September 2001 and 25th February 2002. The horizontal dotted lines represent a loss of 1\%, 5\% and 20\% SR in $H$-band (1.65 $\mu$m).}
\end{figure}

In the top plot of Fig.~\ref{mean_altitude}, we show an example of the derived mean altitude for a profile acquired during the night of the 28th of November 2011. For this given night, the mean altitude changed by $\sim$6 km, which, according to eq. \ref{eq_foc} and \ref{eq_foc2} would correspond to a loss of SR of $\sim$100\% in $H$-band (1.65 $\mu$m). The bottom plot shows how the defocus error changes over different periods of time $\Delta t$, for four sample nights. This plot is computed as an average of the mean altitude differences for a given $\Delta t$. The altitude difference is then converted into a defocus error by using Eq. \ref{eq_foc}. As the figure shows, the larger the time between re-focussing, the greater the variation in the the mean altitude, and thus the more the focus error grows. Note that the dashed lines represent a loss of 1\%, 5\% and 20\% SR in $H$-band (1.65 $\mu$m). Note also the large variation from night to night: one can expect a factor of two in the amplitude of the defocus error. To cope with this issue, LGS-AO systems are usually equipped with an independent focus sensor looking at a natural guide star so as to detect any deviations from the optimal focus, since focus variations on a NGS can only be induced by the atmosphere. The rate at which this sensor must work sets the requirement on the guide star magnitude, hence on the sky coverage. For the examples of Fig.~\ref{mean_altitude}, if we do not want to degrade the SR by more than 5\%, a focus update should be done every $\sim$3 min. If we reduce the requirement to 1\% ($\Delta h$ = 43 m), the focus update should be done every $\sim$30 s. In the current scheme for GeMS, we are using  refocusing rates ranging from 1 second to 5 min depending on the magnitude of the NGS. Good focus measurements are typically obtained for 1s exposure time on NGS with $R$ $<$ 13.0. This means that if we want to keep the focus error down to a loss of 1\% of SR in $H$-band, our current limiting magnitude is about $R$ = 16.7. For fainter stars, the focus error will grow. The current limiting magnitude is above the original requirement of $R$ = 18.5, and thus this limitation in magnitude is impacting sky coverage. This is a known issue of the current hardware in GeMS \citep{Neichel10}. Performance is expected to be improved in the future. \\

\begin{figure}
  \begin{center}
      \includegraphics[width = 0.9\linewidth]{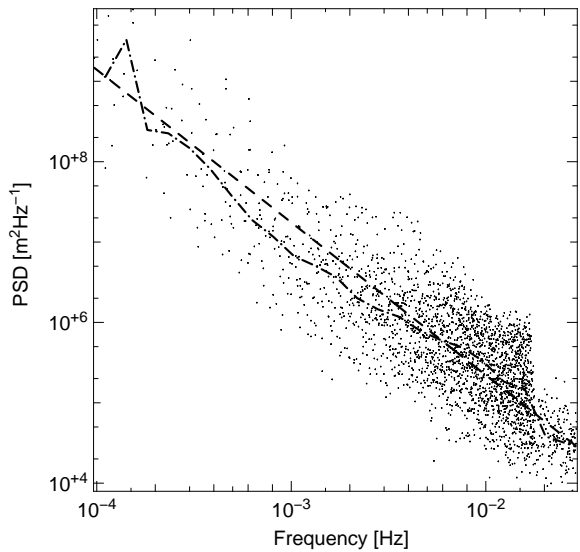} 
  \end{center}
  \caption[] { \label{psd} Power Spectral Density of the mean profile altitude. Dots are for individual data points, the dot-dash line is the average over the different nights and the dashed line is the best fit to the data. }
\end{figure}

To generalize the results presented in Fig. \ref{mean_altitude}, we compute the Power Spectral Density (PSD) of the mean altitude variations for all the nights in our sample. This is shown in Fig. ~\ref{psd}. As found by \citet{Pfrommer10B}, the PSD of the mean altitude variations is well fitted by a model defined by: 
\begin{equation}
\mbox{PSD}(\nu) = \alpha \nu^{\beta}
\end{equation}
where, for our set of data, we find a best fit by using $\alpha$=35 m$^2$Hz$^{-1}$ and $\beta$=-1.9 (dashed line in Fig. \ref{mean_altitude}). This is in good agreement with the results found by \citet{Pfrommer10B} from data acquired in the northern hemisphere, where they derive $\alpha$=30$\pm20$ m$^2$Hz$^{-1}$ and $\beta$=$-1.95\pm0.12$. \\

\subsubsection{Sodium return flux}
\begin{figure}
  \begin{center}
    \begin{tabular}{c}
      \includegraphics[width = 0.9\linewidth]{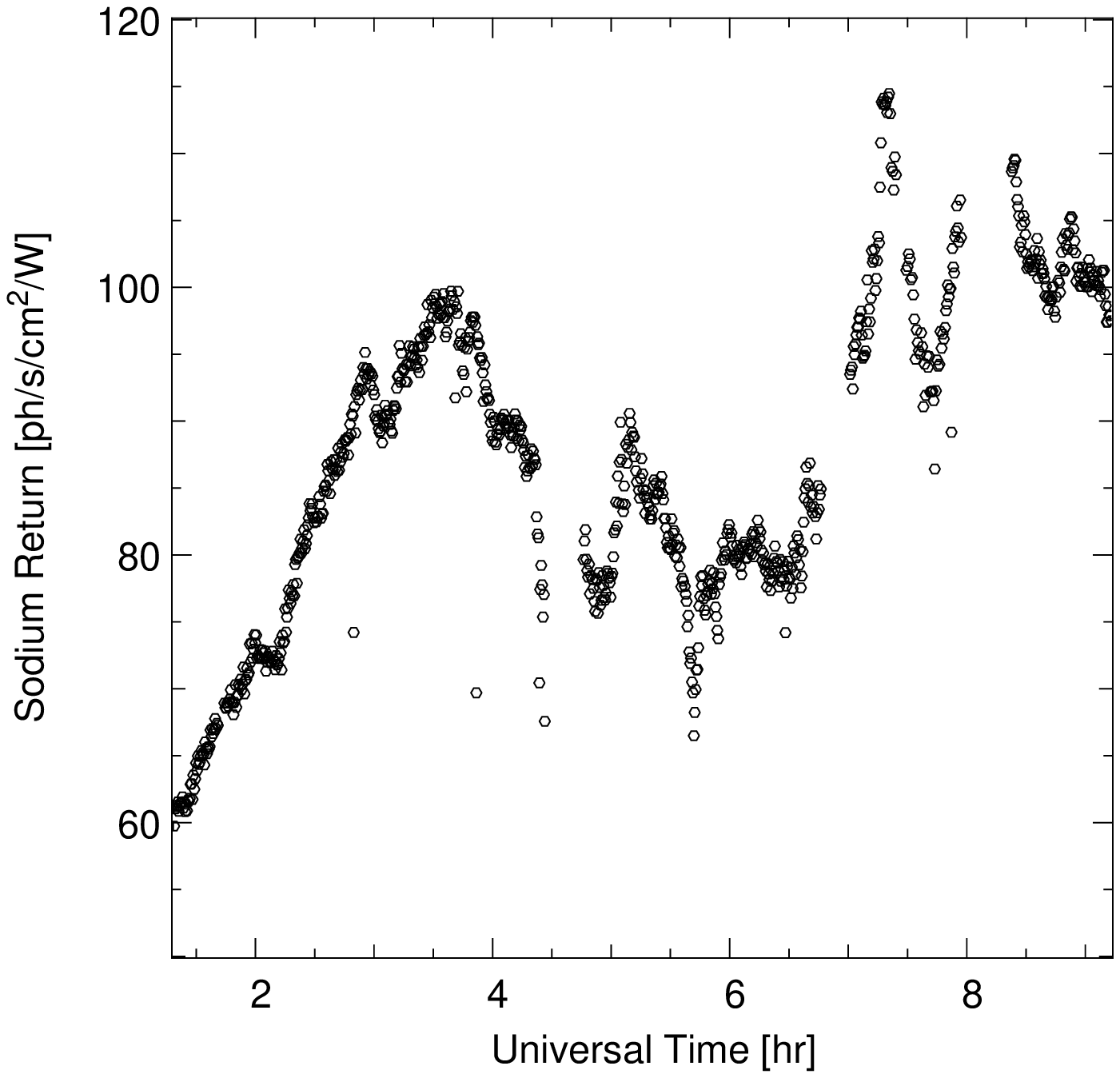} \\
      \includegraphics[width = 0.9\linewidth]{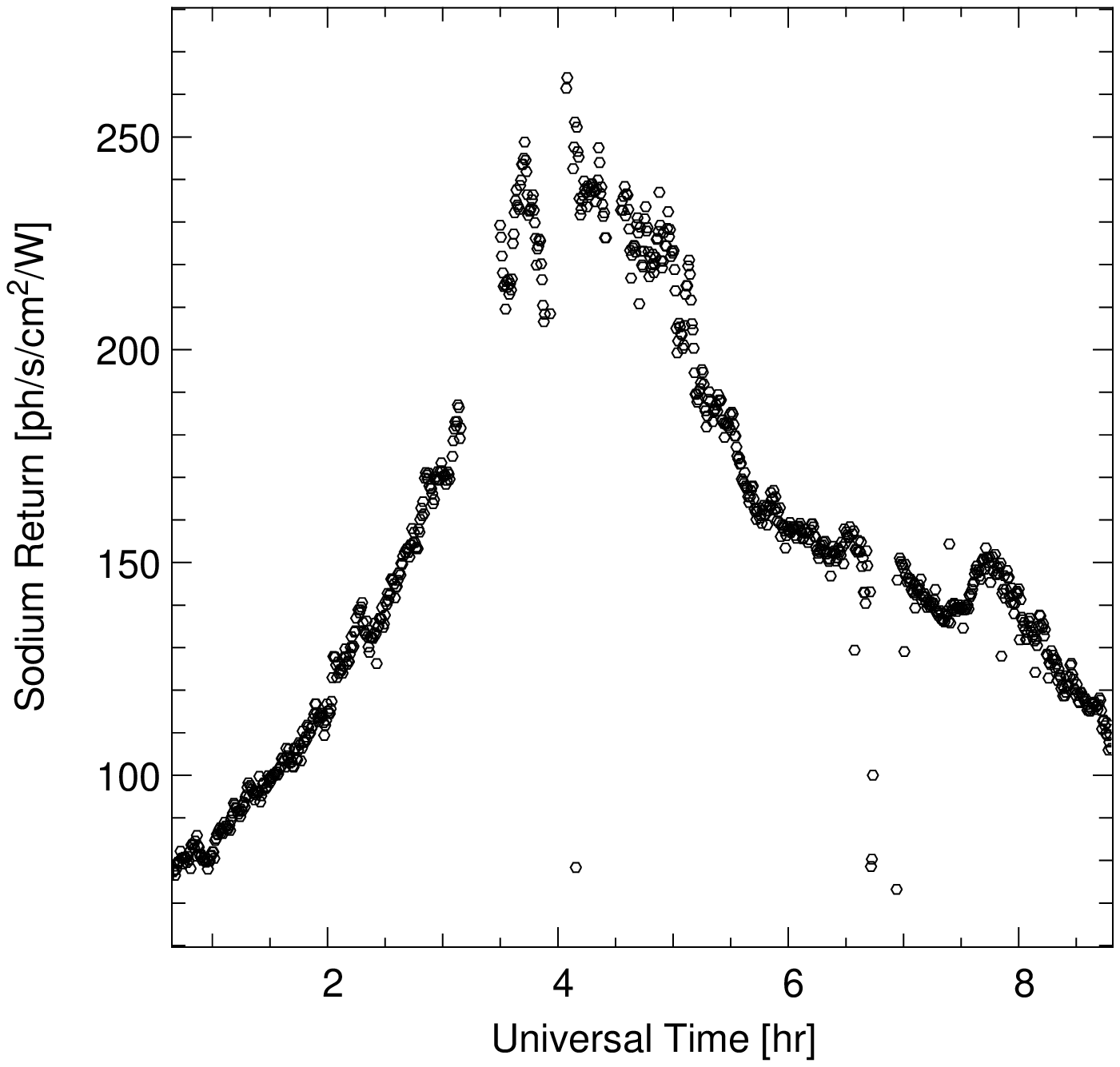}
    \end{tabular}
  \end{center}
  \caption[] { \label{photon_return} Variation of the sodium photon return for two nights: top is February 25th 2002 and bottom is November 28th 2001. The return is given at the sodium layer (atmospheric transmission is not factored in on the way back to the ground).}
\end{figure}

\begin{figure}
  \begin{center}
    \begin{tabular}{c}
      \includegraphics[width = 0.9\linewidth]{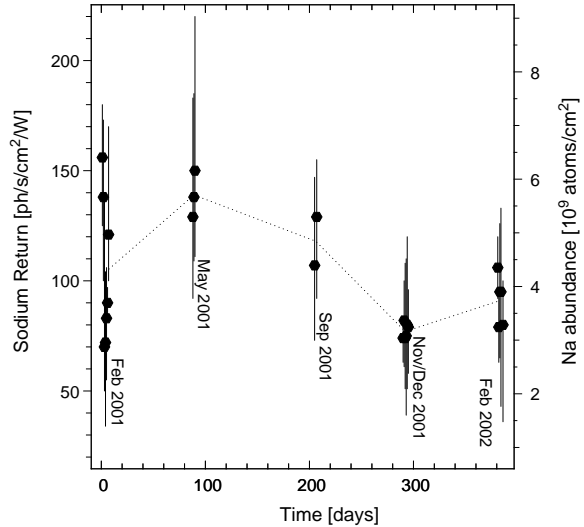}
    \end{tabular}
  \end{center}
  \caption[] { \label{seasonvar} Average photon return and Sodium abundance (in x 10$^9$ atoms/cm$^2$) per night for all the nights that had useable data. The return is given at the sodium layer (atmospheric transmission is not factored in on the way back to the ground). Error bars represent the minimum and maximum values per night.}
\end{figure}

This parameter is critical for LGS AO systems as it is directly related to the flux received by the LGSWFS of the AO system. The sodium return (or brightness of the LGS) can change rapidly during a night, and it is expected to reach its lowest average level during summer time, possibly in November/December due to variations in the sodium abundance. In Fig. \ref{photon_return}, we plot the evolution of the sodium photon return during the night of the 25th of February 2002 and for the night of the 28th of November. One can see that the return can vary by up to a factor of three within the same night. This impacts the AO correction, and should be taken into account by automatic optimization procedures \citep{Neichel10}. The seasonal variation observed during the one year duration of these measurements is shown in Fig \ref{seasonvar} in terms of photons/cm$^2$/s per Watt of projected laser power, given at the sodium layer (atmospheric transmission is not factored in on the way back to the ground) and in terms of sodium column density. The average sodium column density varies by a factor of $\sim$2 between a summer minimum of 3 10$^9$ atoms/cm$^2$ in November/February and a winter maximum of 6 10$^9$ atoms/cm$^2$ in May. These results are in reasonable agreement with sodium abundance values reported in previous studies \citep{Papen96,Michaille01,Moussaoui10A} and confirm that the sodium properties for the Gemini-South site are similar to those of other sites. The impact of the seasonal variations on GeMS performance is discussed in section \ref{sec:gems_return}.

\section{GeMS commissioning}

\subsection{Overview}
GeMS is the first multi-sodium based LGS AO system used for astronomy\citep{Celine08,Celine12,Rigaut12}. The GeMS laser is a CW mode-locked 50 W laser, whose spectro-temporal and spatial format is quite different from the Cerro Tololo laser format, and significantly less efficient in exciting sodium atoms at equal power levels. The Gemini South (GS) laser contains two infra-red (IR) laser lines (respectively 1064 nm and 1319 nm), each created by one oscillator and multiple amplifiers, combined together in a lithium triborate (LBO) non-linear crystal. Each NIR laser is actively mode-locked, resulting in a 77 MHz pulse train with nominal pulse widths on the order of 300-400 ps. The spectral bandwidth of the 589nm laser has been measured to be on the order of 1.5-2.1 GHz. The laser beam is relayed from the output of the Laser system to the input of the Laser Launch Telescope (LLT) located behind the telescope secondary mirror, by a set of mirrors called Beam Transfer Optics (BTO). Five LGSs are produced after splitting the 50 W laser beam into five 10W beams in the BTO. 
The throughput of the BTO has been measured to be $\sim$50\%, and the projected beams are not controlled in polarization yet. Each LGS is seen by a dedicated LGSWFS made of a 16x16 subaperture Shack-Hartmann lenslet array. Each subaperture is sampled by 2x2 pixels (quadcell configuration). The effective laser power projected into the sky is calibrated with respect to the GeMS laser system output power and the output power is monitored at a rate of 1~Hz.  The projected beams are gaussian, with beam diameters on the order of 25 cm at the 1/e$^2$ intensity points. Spot size measured on the sky are ranging between 1.2 and 1.7 arcsec. LGSWFS data has been regularly saved during the GeMS commissioning nights. The photon return is extracted from this data.

\subsection{Data and data reduction}
\label{nlabel}

GeMS started on-sky commissioning in January 2011, at a rate of 5 to 7 nights per month. Data was obtained during the first 5 months of 2011, after which GeMS entered a shutdown phase for engineering upgrades. On-sky operations resumed in November 2011, and continued up to May 2012 again at a rate of one run per month. We do not use the data from January 2011 and February 2011, because at that time the laser spots were not yet optimized, and their FWHM was larger than the LGSWFS field stop, so flux measurements may be biased. From March on, laser spots were of the order of 1.3 to 1.5 arcsec, which is smaller than the field of view of the LGSWFS subapertures of 2.8 arcsec. For all these data points, the laser stabilization loop which is keeping each of the LGS in front of the LGSWFS was closed, so we do not expect flux loses because of coupling with the subaperture FoV. Table \ref{tab:fonts} summarizes the data available for each run. All the data was taken at zenith, in the same conditions, and nights with clouds have been discarded.\\

The data reduction is fairly straightforward in that case. The photon return is measured at the LGSWFS level, by integrating the flux over all the subapertures. The flux is first converted into photons/second/cm$^{2}$ by using the LGSWFS CCD detector gain. Then, the normalization by the laser power projected to the sky is applied. All the data is time-stamped, which facilitates its cross-correlation. The LGSWFS data is acquired at a frame rate ranging from 100 Hz to 800 Hz. However, as the laser power is monitored at a rate of 1~Hz, we average the LGSWFS data over bins of 1 s. Finally, it is important to note that as the flux is measured at the LGSWFS level, it includes all the losses due to transmission of the atmosphere, the telescope itself, and of Canopus (the GeMS AO bench). To translate this flux in terms of photons/cm$^2$/s per Watt of projected laser power at the ground i.e. at the primary mirror of the receiving telescope, one can use the following transmission: T$_{\mbox{tel}}$ $\sim$ 0.8 (measured Gemini South telescope throughput), T$_{\mbox{AO}}$ $\sim$ 0.28 (measured AO bench and LGSWFS throughput at 589 nm), and LGSWFS QE = 0.8 (detector quantum efficiency from constructor data).

\begin{table}
\caption{Summary of data used for the sodium return characterization from GeMS commissioning and average return computed for each of these data sets, as seen by the LGSWFS.} 
\label{tab:fonts}
 \begin{tabular}{@{}ccc}
\hline
Date & Number of nights & Sodium return\\
 &  & (in ph/s/cm$^2$/W)\\
\hline
March 2011 & 3 nights & 5.4 \\
April 2011 & 4 nights & 8.7 \\
May 2011 & 1 night & 9.5 \\
November 2011 & 3 nights & 4.8 \\
December 2011 & 5 nights & 3.5 \\
January 2012 & 4 nights & 4.1 \\
February 2012 & 4nights & 8.2 \\
March 2012 & 5 nights & 6.5 \\
April 2012 & 3 nights & 11.5 \\
May 2012 & 2 nights & 13 \\
\hline 
\end{tabular}
\end{table} 

\subsection{Results}

\subsubsection{Sodium return flux}
\label{sec:gems_return}
\begin{figure}
  \begin{center}
    \begin{tabular}{c}
     \includegraphics[width = 0.9\linewidth]{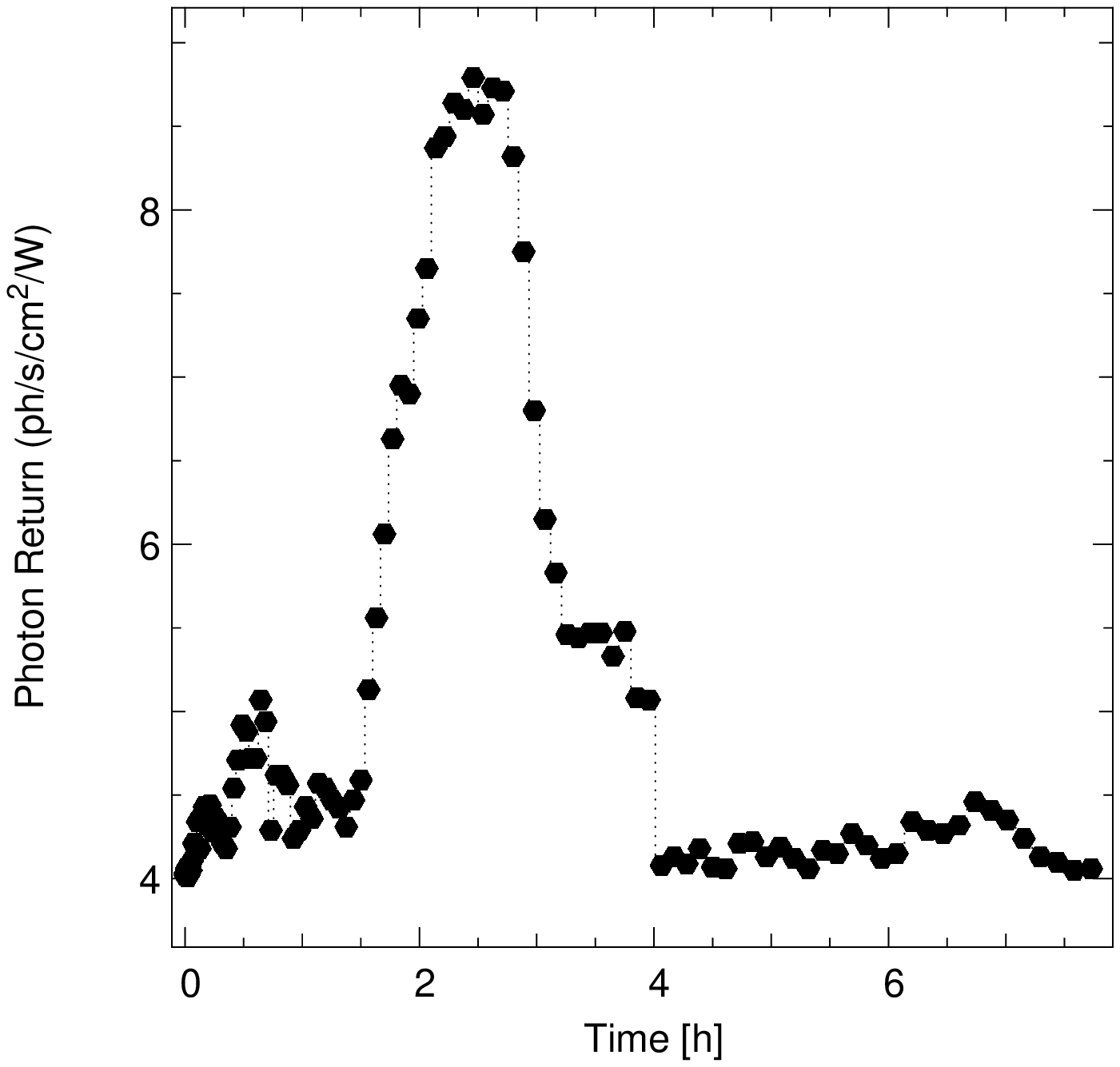} \\
      \includegraphics[width = 0.9\linewidth]{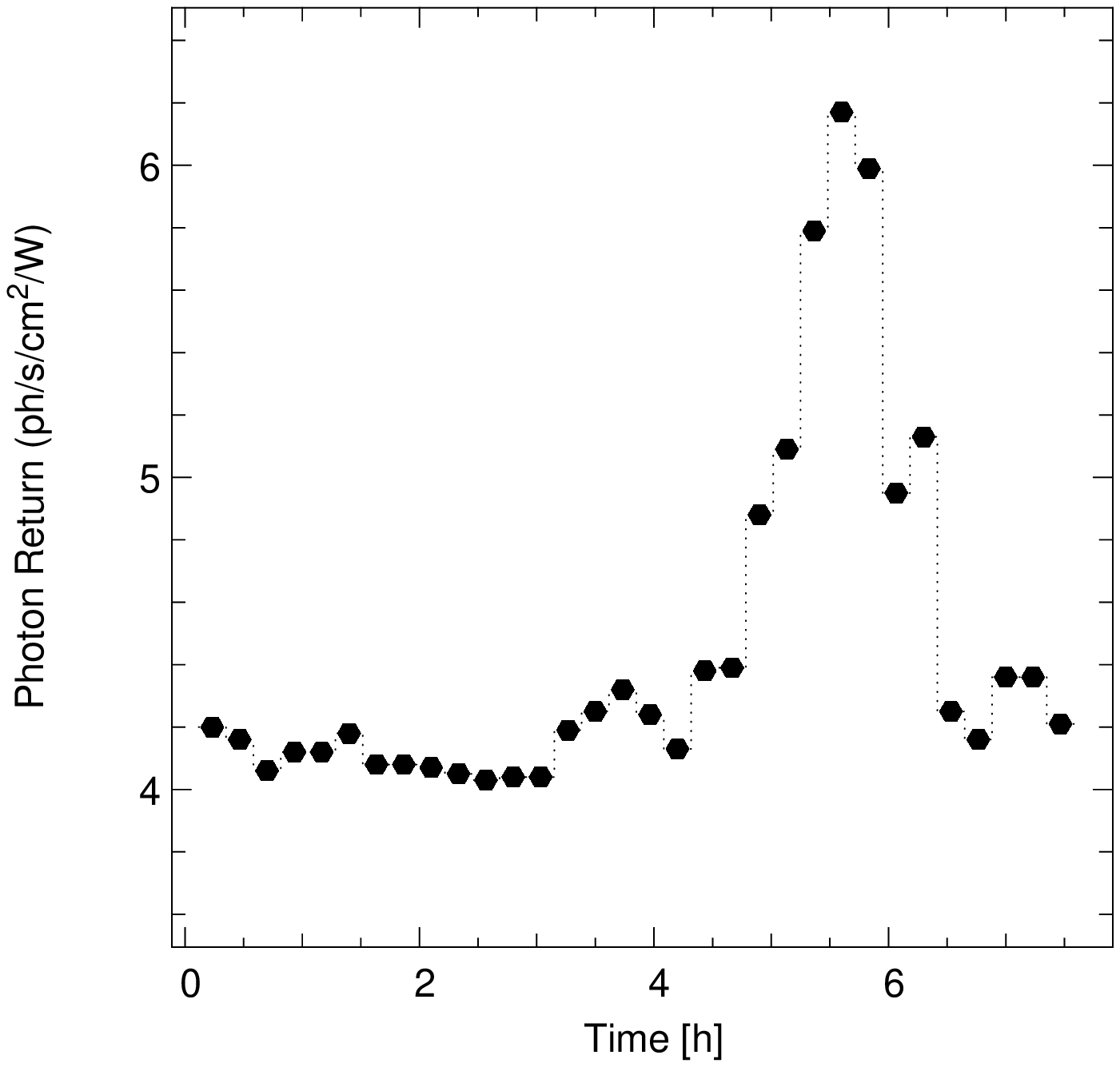}
    \end{tabular}
  \end{center}
  \caption[] { \label{na_return1} Sodium return measured by the LGSWFS for 2 sample nights: 14th of November 2011 (top) and 13th of December 2011 (bottom). }
\end{figure}

Fig.\ref{na_return1} shows the sodium return measured by the LGSWFS for two sample nights: the 14th of November 2011, and the 13th of December 2011. For these two nights, we see that the flux suddenly increases by more than a factor of two, which is most likely due to the presence of sporadics. This is something that is regularly seen, at a rate of once every 2 or 3 nights, and it illustrates the high variability of the LGS brightness that can be expected.
 
\begin{figure}
  \begin{center}
     \includegraphics[width = 0.9\linewidth]{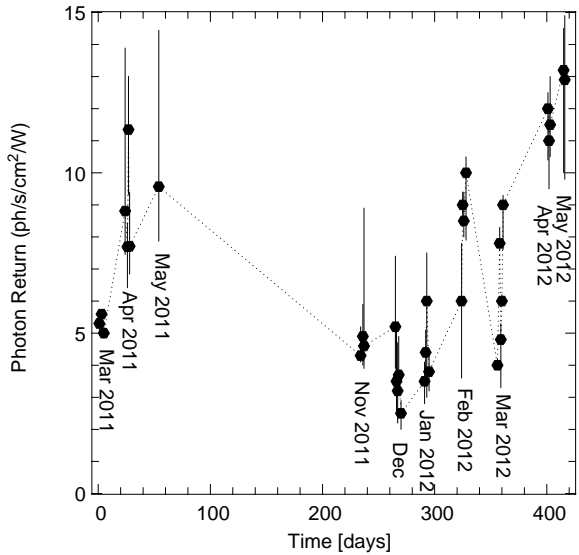}
  \end{center}
  \caption[] { \label{na_return2} Sodium return measured by the LGSWFS for all the data available over more than 1 year of commissioning. Error bars represent the minimum and maximum values per night.}
\end{figure}

\noindent
Fig.~\ref{na_return2} shows the return flux for all the data available. The amplitude variations over a night are represented by the errors bars that show the minimal and maximal return measured during each night. As presented in Fig.~\ref{seasonvar}, we retrieve here the variations due to the sodium season. Lower return is seen in December, while higher return is around May. The difference between low and high season is on the order of a factor of three to four, in good agreement with other studies \citep{Papen96,Moussaoui10A}. This information can be used to optimize the scheduling of laser operations: between May and September laser operations would be facilitated as far as photon return is concerned. Unfortunately, this seasonal variation of sodium abundance is correlated with the bad weather conditions and median seeing variations across the year. This means that when the sodium density is at its maximum, seeing is also at its maximum. This is limiting the use of AO instruments during the Chilean winter and for instance it is planned for GeMS to be shutdown during the months of July and August every year. Such large seasonal variations also directly impact GeMS performance, as the frame rate of the AO loop should be adjusted consequently in order to keep the level of photons measured by the LGSWFS more or less constant. In order to get a reasonable signal to noise ratio on the WFS, we have established that around 35~ph/frame/pixels are needed. If running at 800 Hz, this translates into 10~ph/s/cm$^2$/W at the LGSWFS. Looking at the results of Fig.~\ref{na_return2}, we see that this condition is only reached during the higher sodium season, i.e. around May. During the months of November and December, we have to reduce the AO frame rate, sometimes as low as 100 Hz in order to keep the right SNR at the LGSWFS level.  \\

%

\section{GeMS and Tololo sodium photon return comparison}
This section presents a side-by-side comparison of the sodium photon returns measured in 2001-2002 above Cerro Tololo using a low power, 100 mW-class CW monomode laser (about ~350 mW projected on sky), and in 2011-2012 above Cerro Pach\'on using a high power, 50 W-class CW mode-locked laser (about ~4 W projected on-sky per LGS). Cerro Tololo and Cerro Pach\'on are about 10.5 km away, close enough that the sodium layer properties can be assumed to be the same above these two sites, albeit for measurements made 10 years apart. Table \ref{tabtab} presents a summary of the Tololo and Pach\'on laser power and spectro-temporal formats, which emphasizes the fundamental differences between the two laser probes. It is by now well-known that not all 589 nm lasers are created equal in their interactions with the sodium layer, and that key site characteristics such as the site latitude and the geo-magnetic field strength and its orientation also strongly influence sodium photon return results for a given on-sky laser pointing above that site \citep{Rochester12,Holzlohner10,Miloni98,Michaille01,Moussaoui10A,Celine03,Milonni99}. With this knowledge, we will compare relative sodium abundance variations measured with both lasers, and discuss the relative efficiency of the Gemini South laser format with respect to the Tololo laser format.

\begin{table}
\caption{Summary of the main characteristics for the Tololo and GeMS laser.} 
\label{tabtab}
 \begin{tabular}{@{}ccc}
\hline
Description & Tololo Laser & GeMS Laser \\
\hline
Average Power & 150-350 mW & 35-45 W \\
Type & CW & Pulsed \\
Pulse repetition &  & 77 MHz \\
Pulse length &  & 300-400 ps \\
Spectral bandwidth & 0.5 MHz & 1.5-2.1 GHz \\
\textbf{Polarization} & \textbf{Linear} & \textbf{Elliptical (not controlled)} \\
\hline 
\end{tabular}
\end{table} 

Fig. \ref{seasonvar} presents sodium photon return results obtained on Tololo both in terms of photons/cm$^2$/s/W of projected laser power, at the sodium layer, and in terms of sodium column density. Fig. \ref{na_return2} presents sodium photon return results obtained on Pach\'on only in terms of photons/cm$^2$/s/W of projected laser power, at the GeMS LGSWFS. At this time there is no direct sodium abundance measurement concurrent to our data to enable accurate calibration of the sodium column density corresponding to the GeMS sodium photon return data. This is unfortunate since it would have made possible to compare not only relative but also absolute variations in sodium abundance. As it is, the conversion factor between sodium photon return from the GeMS laser and the corresponding sodium column density can only be estimated using the relatively rough assumptions presented below. 

In order to compare the Tololo and GeMS sodium photon returns, we choose to express both of them in terms of photons/cm$^2$/s per Watt of projected laser power at the ground i.e. at the primary mirror of the receiving telescope. For the GeMS data, we use the conversion factors introduced in Sect. \ref{nlabel}. For the Tololo data, we use the atmospheric transmission derived in Sect. \ref{ddr}. Nightly sodium photon return averages obtained in 2001-2002 at Tololo (Fig. \ref{seasonvar}) thus translate into equivalent returns of $\sim$60 to $\sim$130 photons/cm$^2$/s/W at the ground, while nightly sodium photon return averages measured with GeMS in 2011-2012 (Fig. \ref{na_return2}) are in the 13 to 70 photons/cm$^2$/s/W range at the ground. 

Fig. \ref{seasonvar} shows that the monthly November/December 2001 average for sodium abundance is on the order of $\sim$3.5 10$^9$ atoms/cm$^2$, corresponding to $\sim$65 photons/cm$^2$/s/W at the ground. Combining the monthly averages provided in Table \ref{tab:fonts} for the months of November and December 2011 yields a weighted November/December 2011 average of $\sim$10.5 photons/cm$^2$/s/W at the ground. Assuming that the low season, November/December monthly average value in both data sets corresponds to the same low value for sodium column density monthly averages, this yields a conversion factor of about 3 between the Tololo CW laser format and the GeMS laser format. Doing the same thing for high season values in May 2001 (Fig. \ref{seasonvar}) and May 2011/2012 (Table \ref{tab:fonts}) respectively yields $\sim$116 photons/cm$^2$/s/W on the ground at Tololo and $\sim$66 photons/cm$^2$/s/W on the ground at Pach\'on, resulting in a smaller conversion factor of $\sim$1.8. The high season approach is deemed to be much less reliable based on what is known of typical sodium abundance variation behavior, where sodium abundance can spike at times due to high rates of sporadics over a set of nights, whereas low sodium column density values typically remain of the same order of magnitude. The real conversion factor between the two laser formats therefore lies somewhere between 1.8 and 3, and is believed to be closer to 3. Not surprisingly, the Tololo monomode CW laser appears to be several times more efficient in exciting sodium atoms than the Gemini South laser by a large fraction. How large this factor really is remains to be determined with higher accuracy. 

The sodium photon return results for the Tololo laser can also be compared to table 2 presented in \citep{Telle06}. Telle's Fasor laser is a high power (50W) CW monomode laser for which results extrapolated to low (``zero'') power basically apply to the Tololo laser format. Telle reports that, based on extrapolating the Fasor on-sky measurements, such a laser format provides 150 (resp. 140, 185, 115) photons/cm$^2$/s/W ``at zero power'' of projected Fasor power at the ground at the Starfire Optical Range in New Mexico, USA, on October 12 (resp. October 14, November 16, and December 22), 2005, for an estimated sodium column density of 7.5 (resp. 7.0, 8.4, and 5.9) 10$^9$ atoms/cm$^2$. Scaling these values down to our assumed, Tololo low sodium column density value of 3.5 10$^9$ atoms/cm$^2$ yields sodium photon return values of 70 (resp. 70, 77, and 68) photons/cm$^2$/s/W of projected Tololo laser power at the ground, in rather good agreement with our estimated value of $\sim$65 photons/cm$^2$/s/W. This comparison gives us reasonable confidence in the Tololo results presented in section 2 and in particular in the sodium column density values we have derived from them.

Sodium abundance measurements have been performed on Cerro Pach\'on by the neighboring Andes LIDAR Observatory (ALO, the University of Illinois sodium LIDAR experiment, see \citet{ALO09}) during various observation runs since 2009. Unfortunately no ALO data is available concurrent with GeMS sodium photon return data which would have permitted a direct calibration of the sodium abundance based on GeMS results. Some data is available in between runs though, which we can use to determine the validity of our earlier estimates of how much more efficient the Tololo laser is with respect to the GeMS laser. ALO sodium abundance data \citep{ALO12} was obtained over the nights of January 28-February 2, 2012 (average sodium column density $\sim$3.9 10$^9$ atoms/cm$^2$) and March 20-23, 2012 (average sodium column density $\sim$4.9 10$^9$ atoms/cm$^2$), in between the GeMS January 7-12, February 10-13, and March 10-14, 2012 nights used to derive the sodium photon return results presented in Table 1. Table \ref{ctable} presents a summary of the relevant ALO, Tololo and GeMS data which can be used to compare actual vs. extrapolated GeMS sodium photon returns in the January-March 2012 period. Expected sodium photon returns are calculated for the Tololo and GeMS laser formats assuming a conversion factor of 3 between these two formats. Measured and extrapolated GeMS results in the January-March 2012 period are qualitatively consistent, reinforcing our belief that a factor of 3 is probably close to the mark.


\begin{table*}
\caption{Comparison between Tololo and GeMS sodium photon results using ALO sodium abundance data to extrapolate results between data sets. Extrapolated results are indicated in bold italics.} 
\label{ctable}
 \begin{tabular}{@{}ccc|cc|cc}
\hline
\multicolumn{3}{c}{Data measurement period} & \multicolumn{2}{c}{Sodium abundance} & \multicolumn{2}{c}{Photon return on ground} \\
\multicolumn{3}{c}{} & \multicolumn{2}{c}{(atoms/cm$^2$)} & \multicolumn{2}{c}{(ph/cm$2$/s/W)} \\
\hline
Year & Month & Day & ALO & Tololo & Tololo & GeMS \\
\hline
2001 & Feb & 11-20 & & 4.3E+09 & 89.3 & \\
2002 & Feb/March & 23-28,01 & & 3.8E+09 & 76.5 & \\
2012 & Jan & 7-8, 10-12 & & & & 22.9 \\
2012 & Jan/Feb & 28, 30-31, 01-02 & 3.9E+09 & & \textbf{\textit{80.0}} & \textbf{\textit{26.7}} \\
2012 & Feb & 10-13 & & & & 45.9 \\
2012 & March & 10,12-14 & & & & 36.2 \\
2012 & March & 20-23 & 4.9E+09 & & \textbf{\textit{101.9}} & \textbf{\textit{34.0}} \\
\hline 
\end{tabular}
\end{table*}

We are hopeful that, in the future, closer coordination with the Andes LIDAR Observatory will permit concurrent GeMS laser and LIDAR laser propagation above Cerro Pach\'on. Direct calibration of the sodium abundance at the same time as GeMS sodium photon return data is gathered will likely prompt a review or at least a refinement of the conclusions presented here.

\section{Conclusion}
We have presented results on the sodium photon return and characteristics of LGS created above Cerro Pach\'on and Cerro Tololo based on two sets of data. Both were gathered over more than a year of observations and have been used to compute the sodium photon return per Watt of projected laser, the sodium profiles and the variations of the sodium layer mean altitude. The results show that: 
\begin{itemize}
\item A LGS exhibits a large variety of fluctuations in shape and brightness. These changes can appear on very short (few seconds) or seasonal time scales.
\item Average sodium equivalent width are on the order of $\sim$10 km, which leads to a spot elongation of $\sim$1 arcsec for an 8 m telescope using an on-axis launch configuration.
\item A small correlation of the sodium equivalent width with seasons is detected: the sodium layer is thinner by 1km around September.
\item The PSD of the sodium mean altitude variation is well fitted by a model defined by $\mbox{PSD}(\nu) = \alpha \nu^{\beta}$. We derived $\alpha$=35 m$^2$Hz$^{-1}$ and $\beta$=-1.9 in good agreement with previous data published in Northern hemisphere.
\item A refocusing rate of $<\sim$30 s is required for GeMS.
\item The sodium photon return shows a large variability over a year, with seasonal fluctuations on the order of 3 to 4. 
\item Nightly fluctuations of photon return by a factor of two are regularly seen.
\item Comparing sodium photon returns obtained with two different lasers clearly illustrates the impact of the laser format on coupling efficiency with the sodium layer. For the two laser format described in Tab. \ref{tabtab} and used in this study, we measure a difference in efficiency by a factor of $\sim$3.
\end{itemize}


\noindent
\textbf{Acknowledgments}\\

\noindent
J. Callingham recognises the support of the Australian Astronomical Observatory (AAO) via the Australian Undergraduate Gemini Summer Studentship (AGUSS). Authors want to acknowledge Alan Liu, Gary Swenson, Fabio Vargas and Anthony Mangognia for kindly providing the ALO data. The Andes Lidar Observatory is supported by NSF and University of Illinois. The Gemini Observatory is operated by the Association of Universities for Research in Astronomy, Inc., under a cooperative agreement with the NSF on behalf of the Gemini partnership: the National Science Foundation (United States), the Science and Technology Facilities Council (United Kingdom), the National Research Council (Canada), CONICYT (Chile), the Australian Research Council (Australia), 
Minist\'{e}rio da Ci\^{e}ncia e Tecnologia (Brazil) and Ministerio de Ciencia, Tecnolog\'{i}a e Innovaci\'{o}n Productiva (Argentina).

\label{lastpage}

\end{document}